\documentclass[12pt,preprint]{aastex}

\shorttitle{The GUSBAD Catalog of Gamma-Ray Bursts}
\shortauthors{Schmidt}

\begin{document}

\title{The GUSBAD Catalog of Gamma-Ray Bursts}

\author{Maarten Schmidt}
\affil{California Institue of Technology, Pasadena, CA 91125}

\begin{abstract}
The GUSBAD catalog of gamma-ray bursts (GRBs) is based on 
archival BATSE DISCLA data
covering the full 9.1 years of the Compton Gamma Ray Observatory mission.
The catalog contains 2204 GRBs, including 589 bursts not listed in the
Current BATSE Burst Catalog. The GUSBAD catalog is uniform in the sense
that the detection criteria are the same throughout and that the
properties given in the catalog are available for every burst. The 
detection and the derivation of the properties of the GRBs were carried
out automatically. This makes the GUSBAD catalog especially suitable
for statistical work and simulations, such as used in the derivation of
$V/V_{\rm max}$. We briefly touch upon a potential problem in defining
a GRB duration that is physically meaningful.

\end{abstract}

\keywords{cosmology: observations --- gamma rays: bursts}

\section{Introduction}
The Burst and Transient Source Experiment (BATSE) \citep{fis89}
on board the {\it Compton Gamma Ray Observatory (CGRO)} has been
very successful in detecting gamma-ray bursts (GRBs).  The data 
have been published in a succession of catalogs, the most recent 
of which is the Current BATSE Burst Catalog\footnote{Available at 
http://www.batse.msfc.nasa.gov/batse/grb/.} ('the BATSE catalog').
The 2702 GRBs in the BATSE catalog are the result of the full
mission of {\it CGRO} from 1991 April 19 to 2000 May 26.

Burst detections are based on counts recorded by eight Large Area
Detectors (LADs) located at the corners of the spacecraft
\citep{fis89}. Counts are collected in four energy channels
($20-50$, $50-100$, $100-300$ and $>300$ keV) on time scales of
64, 256, and 1024 ms. All GRBs in the BATSE 
catalog are based on an on-board trigger mechanism, which acted 
when certain conditions were fulfilled. Usually these required
that the LAD counts in the energy range $50-300$ keV exceeded the 
background by at least $5.5 \sigma$ on a time scale of 64, 256 
or 1024 ms in at least two of the eight BATSE detectors. 
The background for each detector was averaged over 17.408 s, 
immediately preceding the burst and was recomputed every 17.408 s 
\citep{fis89}. The trigger
mechanism was disabled for up to 90 minutes following a burst
detection to allow telemetering burst data to the ground. It also 
was disabled during passage through regions with a high density 
of atmospheric particle precipitation events \citep{fis94}. 
The BATSE catalog gives for each GRB the time of detection, celestial 
coordinates, maximum and minimum count rates, peak flux, fluence
and durations whenever available.

For statistical studies the BATSE catalog has some serious drawbacks. 
The maximum and minimum count rates, needed to derive $V/V_{\rm max}$,
are available for only $\sim$ 49\% of the 2702 GRBs. Also, the catalog
is not uniform in the sense that the trigger criteria were changed
many times through the mission; the standard parameters mentioned above 
were in effect for $\sim$ 55\% of the mission duration. 

A catalog of GRBs without these drawbacks can be constructed from
archival data produced by BATSE. For this purpose we have used the 
DISCLA data which provide a continuous record of the LAD 
counts in the four energy channels on a time scale of 1024 ms
for each of the eight BATSE detectors.  These data allow the
{\it a posteriori} detection of GRBs with a software trigger in 
a manner similar to that executed by the on-board trigger mechanism.
The availability of only the 1024 ms timescale means that the search
wil be essentially limited to bursts with a duration of more than
$1-2$ s. There are distinct advantages in using a software trigger 
on the continuous data stream. One can experiment with the derivation 
of the background or can repeat the search for bursts with
different detection criteria.  We have carried out such a search 
resulting in the GUSBAD catalog (Gamma-ray bursts Uniformly Selected 
from BATSE Archival Data) \citep{sch03}\footnote{Available at 
http://www.astro.caltech.edu/$\sim$mxs/grb/GUSBAD/.} The GUSBAD catalog 
supercedes earlier work on DISCLA data reported in \citet{sch99a,sch99b}.

In order to allow simulations, such as used in the derivation 
of $V/V_{\rm max}$, we were guided by the following precepts in the
search for GRBs. The detection and the derivation of burst events
should be carried out automatically. We search for bursts only 
at times when all properties of a burst event can be derived. 
Also, we treat strong bursts no differently from weak bursts. 
We worked independently from other catalogs, such as those of
\citet{kom97} and \citet{ste01}, which were under development while
this work was in progress; however, in the classification
procedure (see Sec. 3) we did use the listing of any GUSBAD burst event 
in the BATSE catalog as confirmation that the event was a cosmic GRB.

We describe in Sec. 2 the treatment of the DISCLA data, the definition
of the trigger mechanism and the derivation of the celestial coordinates 
of burst events. In Sec. 3 we discuss the
classification of the triggers required to separate the cosmic GRBs
from other types of events. Sec. 4 covers the derivation of exposure 
and effective limiting peak flux and Sec. 5 the simulation producing
the Euclidean value of $V/V_{\rm max}$. The GUSBAD
catalog is described in Sec. 6. The discussion in Sec. 7 includes
comments on the problem of defining a robust GRB duration.

\section{Burst Events from DISCLA Data}

\subsection{DISCLA Data} 

DISCLA data provide every 1024 ms counts in the four energy channels
for each of the eight
BATSE LAD detectors \citep{fis89} and every 2048 ms information
about the orientation and geocentric coordinates of {\it CGRO}. We used
for the time period TJD $8365-10528$ a tape copy of DISCLA data made
available by T. Prince,
and for TJD $10529-11690$ data obtained from the High Energy
Astrophysics Science Archive Research Center (HEASARC).

\subsection{Trigger Definition}

For the detection of a burst we use a software trigger requiring that
the counts in the $50-300$ keV energy range exceed the background by
at least $5.0 \sigma _B$ in two or more of the eight BATSE detectors,
where $\sigma _B$ is the standard deviation of the background counts.
Let $C_d(k)$ be the measured number of counts in 1024 ms time bin $k$
for detector $d$. To estimate the number of background counts $B_d(k)$,
we average the counts in time bins $k-n_p-n_b$, ..., $k-n_p-1$ to produce 
$B_{d1}$, and in time bins $k+n_f+1$, ..., $k+n_f+n_b$ to get $B_{d2}$. 
We adopt $n_b=17$ as was done for the on-board BATSE trigger. We also 
use $n_p=20$ and $n_f=225$ and derive $B_d(k)$ from a linear interpolation
between $B_{d1}$ and $B_{d2}$, see Figure 1. The signal-to-noise ratio
of the excess counts is $S_d(k) = [C_d(k)-B_d(k)]/[B_d(k)^{1/2}$; if it
is larger than 5.0 in two detectors, we record the onset of a
burst event in time bin $k_{\rm trig}$.

The interval $n_p$ between the first background interval and the test 
bin was introduced to allow detection of slowly rising bursts which may 
have escaped detection with the BATSE trigger \citep{hig96}. In setting
the $n_f$ value, we essentially assume that the duration of the GRB is 
less than 230.4 s. The effect of some burst signal in the second 
background interval usually will be minor given its low weight in the 
interpolated background. Some very long bursts required special 
treatment, see Sec. 3. 

To make sure that all properties can be derived for each recorded 
burst, we limit the search to bins in which the counts are recorded and
uncontaminated from $k-n_f-n_b$ to $k+n_p+n_b$. Interruptions were caused
by high voltage switch-off for the South Atlantic Anomaly; poor data
were identified by checking for instances where several detectors recorded 
a zero or constant count. We also excluded appropriate time intervals 
around checksum errors reported in DISCLA data, which turned out to 
mimick short bursts. Ultimately, we identified a total of 199,964 time 
windows of different lengths which were excluded from the search. 

In an early search for GRBs based on the period TJD 8365-10528, we found 
strong concentrations of triggered events recorded in geographical
areas over W. Australia, Texas and an area bordering the South Atlantic
Anomaly \citep{sch99a}. To avoid searching for cosmic GRBs in such a high
density of non-cosmic triggers, we established geographical exclusion 
regions around the areas of highest density. This reduced the total
number of triggers by $\sim$40\%.

\subsection{Celestial Coordinates}

The burst events detected are composed of many different sources besides
cosmic GRBs, see Sec. 3. In the classification procedure required to find 
the nature of the events, celestial coordinates play an important role. 
We discuss here the derivation of positions on the assumption that the 
burst event is a point source. We start by setting up a grid of 
$\sim$ 40,000 positions, separated by $\sim 1\degr$ in a equatorial
coordinate system anchored on the satellite. Using the BATSE DRM 
generator code supplied by J. Brainerd, we derive at each 
of these positions the effective cross-section for each detector,
ignoring the scattered radiation part of the DRM.
We use a Band spectrum \citep{ban93} with $\alpha = -1.0$,
$\beta = -2.0$ and $E_0 = 200$ keV. 
The position of a burst event is derived by finding in
which of the 40,000 grid positions the observed counts produce the
largest amplitude for the burst. Our procedure involves the use of 
all eight detectors. We determine the positions from each time bin
in which the burst is exceeding the minimum detectable flux. 
The ultimate position is derived from the sum of the counts
in all the time bins with reasonably concordant positions; this part
does include an iteration which accounts for the radiation off the
Earth's atmosphere and the satellite. The specific steps are given below.

Based on the background (by linear interpolation between $B_{d1}$ and
$B_{d2}$) used at the time of trigger, derive the net 
burst counts in each detector in 225 time bins starting with 
$k_{\rm trig}$. For each of the 40,000 grid positions, 
carry out a least-squares solution to derive a photon flux from 
the net counts in the eight detectors; the largest derived flux
sets the position of the trigger event. Use this flux $f_{\rm trig}$
and the signal-to-noise ratio in the second brightest illuminated 
detector at trigger to derive the minimum detectable flux $f_{\rm min}$ 
corresponding to a signal-to-noise ratio of 5.0. Derive the positions
independently for each of the 225 time 
bins in which the flux $f>f_{\rm min}$.
Starting at $k_{\rm trig}$ successively monitor the average position and 
reject positions that deviate by more than $20\degr$ from the average.
For the bins with accepted positions, sum the net burst counts 
accumulated during the burst. Using the summed net counts, derive the 
satellite-anchored celestial coordinates.
Correct the net counts for scattered radiation, 
re-derive the celestial coordinates as before and iterate. From the
final position so derived and the positions obtained from each 
contributing time bin, derive the standard deviation of one position, 
and the apparent angular rate of motion across the sky.

\section{Classification of Events}

The search for burst events covered a total of around 250 million
time bins, each of which was tested for the presence of a burst.
Using the procedures described above, we found 6236 burst events.
Figure 2 shows the equatorial coordinates of all the triggers.
The figure shows four types of events. Most prominent are CygX-1
and GRO J0422+32 ( = Nova Persei 1992) in the N. hemisphere. 
The sun exhibits itself through solar flares along the ecliptic. 
A fourth component appears to be more or less isotropic.

The outburst of GRO J0422+32 affords an opportunity to evaluate 
the accuracy of our burst positions. From TJD $\sim 8840-8900$
the X-ray nova was active, producing a total of $\sim$400 bursts.
At the peak of its activity, we recorded 165 burst events in four 
days, from TJD 8841-44. Three of these events were located at 
distances of 58\degr, 63\degr and 152\degr, respectively, from the nova.
The other 162 were all within 20\degr and clearly associated
with the nova, see Figure 3. The standard deviation from the mean
is $\pm 7.5\degr$. The average photon flux of these events was 
$\sim 0.26$ ph cm$^{-2}$ s$^{-1}$, i.e. similar to the weakest GRBs
in the GUSBAD catalog. 

Plots of the rate of burst events versus time for positions close
to the sun, CygX-1 and GRO J0422+32 show well defined periods of
activity. We eliminated as GRB candidates all events within 17\degr 
of these sources while they were active. Plots of the hardness ratio 
ch 1/(ch 2 + ch 3) versus either angular distance to the sun or
altitude of the sun showed that there were still a number of soft events
assocated with the sun. A correlation of triggers with
known solar flares also produced some positive results. As for
CygX-1 during its less active periods, a number of positive jumps
in the counts were seen corresponding to its rising above the horizon,
and many cases were found where the counts showed fast oscillations 
apparently associated with CygX-1 based on the positions.
Eventually, we eliminated as GRB candidates over 1500 triggers which 
were solar flares or close to the sun, upwards of 760 triggers close 
to or associated with CygX-1 and over 400 triggers near GRO J0422+32. 

At this stage we correlated all remaining events with the BATSE
catalog. All events with onsets within $(-15,+30)$ s from that of
a BATSE GRB and a position difference less than 30\degr were
considered confirmed as cosmic GRBs. The decision whether this 
also applied to the triggers with larger differences in trigger time 
(up to 230 s) or position was mostly based on the time profile.
This excercise showed that 1615 burst events were present in the
BATSE catalog and therefore confirmed as GRBs.

For the $\sim$2100 remaining triggers, we inspected the time profiles
in each of the eight detectors over 700 s around the trigger time,
in some cases in the two brightest illuminated detectors 
over an interval of 12000 s. At this stage, we were guided by
descriptions of magnetospheric events \citep{fis92,hor92} which
caused many of the remaining bursts. In evaluating each event,
besides the time profile we also paid attention to
the values of $\chi^2$ and the rms deviation of the positions
for each time bin (see Sec. 2.3), the apparent angular motion
of the burst event (which for a GRB should be zero), and the
altitude of the event (the horizon being at $-18\degr$). 
This excercise resulted in the rejection of around $\sim 1300$
triggers. 

In addition to six detections of SGR $1806-20$, we have some 50 
triggers from a variable near $285+10$ from TJD $10959-11145$ and 
around 40 near $250-55$ from TJD $10981-11026$. We also checked for 
the presence of variables based on a list provided by B. Harmon,
but found no further ones. 

Based on a review of the surviving GRB candidate triggers, 19 GRBs 
were found to have been detected twice and one GRB three times.
Also, in six cases the derivation of positions was affected by
the occurrence of a second burst, by saturation or by unexpected 
bad data. These cases were all handled individually.
We finally ended up with 589 GRBs that are not listed in the
BATSE catalog. Together with the 1615 GRBs that are in the BATSE
catalog, we have a total of 2204 GRBs in the GUSBAD catalog.

\section{Exposure and Limiting Flux}

We investigate the total exposure and limiting photon flux of the
GUSBAD catalog by setting up a $10 \times 10\degr$ grid of 412
sky positions. Every 100 s during the entire mission, we used our
detection algorithm to derive the background in the second brightest
illuminated detector at each of the 412 positions and with the
DRM generator (ignoring scattered radiation) obtained the 
limiting photon flux corresponding to a signal-to-noise ratio of 5.0. 
The effect of scattered radiation off the Earth atmosphere and the
space craft was similarly explored once every ten days during the mission.
We averaged the results over all sky positions and accounted for 
the rejection of triggers near the sun, CygX-1 and GRO J0422+32 
when active, as well as Earth blockage. The total exposure time was
$1.0052 \times 10^8$ s or 3.185 years. It corresponds to the total
time during which the GRB search was effectively carried out over
the entire celestial sphere. Figure 4 shows the distribution
of exposure with limiting photon flux, both with and without the
effect of scattered radiation. 

For statistical work such as deriving the $V/V_{\rm max}$ value, it may
be sufficient to use an effective photon limit that yields the same
number of sources over the total exposure time as does a proper
evaluation using the distribution of photon limits. For values of
the integral source count slope in the range $-0.5$ to $-1.0$ 
applicable to the faintest GRBs in the GUSBAD catalog, the effective
limiting photon flux is 0.25 ph cm$^{-2}$ s$^{-1}$. If this limit is 
used for statistical purposes, all GRBs listed in the catalog should 
be included.

The annual rate of GRBs averaged over the full mission but corrected
for Earth occultation is $2204/3.185 = 692$ y$^{-1}$. The integral 
GRB source counts $N(>P)$ as a function of peak flux $P$ are shown in
Figure 5. At the bright end the logarithmic slope is close to the
value $-3/2$ consistent with a uniform space distribution in 
Euclidean space. The effect of the decrease of exposure time for 
$P < 0.3$ ph cm$^{-2}$ s$^{-1}$ is clearly seen. 

\section{Deriving $V/V_{\rm max}$}

For extragalactic objects, the Euclidean 
value of $<V/V_{\rm max}>$ is a cosmological distance indicator \citep{sch01}. 
The value of $V/V_{\rm max}$ for a GRB is usually derived from the peak burst
count rate $C_{\rm max}$ and the limiting detection rate $C_{\rm min}$ as
$(C_{\rm max}/C_{\rm min})^{-3/2}$. This is not stricly correct as we 
shall see.

We derive the value of $V/V_{\rm max}$ of an individual GRB by a simulation.
The simulation is best visualized as an excercise in which the distance 
of the burst is incrementally increased in Euclidean space until it 
becomes undetectable. The full time profile of the burst is 
reduced by a factor corresponding to the increased distance
and then added back to the original interpolated background (see Fig.
1). The detection algorithm is employed to search for the reduced burst.
If it is detected, the process - including the full search - is repeated
until the burst is not detected anymore. If the burst is lost when
the distance has been increased by a factor $f$ then $V/V_{\rm max} = f^{-3}$.
The 2204 GRBs in the GUSBAD catalog have $<V/V_{\rm max}> = 0.346 \pm 0.006$. 

In the process of analyzing the reduced burst profiles, the trigger 
may occur at later times depending on the detailed profile. (If this
causes the second background stretch to suffer from contaminated or 
missing data, it is kept at its location during the original
detection.) If so, the
two background stretches defined in Fig. 1 move forward and the first
background stretch may contain burst signal. This reduces the
amplitude of the reduced burst and increases the background.
Hence $V/V_{\rm max}$ derived from the simulation is larger than 
$(C_{\rm max}/C_{\rm min})^{-3/2}$. This is confirmed in the GUSBAD
catalog which yields $<(C_{\rm max}/C_{\rm min})^{-3/2}> = 0.335$. 
In the BATSE catalog, only 884 of the 1144 GRBs with 
$C_{\rm max}/C_{\rm min} \ge 1.0$ on a timescale of 1024 ms can 
be used for a comparison: the others fall at times when the 
on-board trigger used parameters other than the standard mentioned 
in Sec. 1. They yield  
$<(C_{\rm max}/C_{\rm min})^{-3/2}> = 0.311 \pm 0.009$.

\section{The GUSBAD Catalog}

The GUSBAD catalog is available on the World Wide Web
\citep{sch03}\footnote{Available at 
http://www.astro.caltech.edu/$\sim$mxs/grb/GUSBAD/.}. 
The designation of GRBs in the catalog is GUSBAD YYMMDD.ddd where
Y, M, D are integral year, month and day numbers, respectively,  and 
.ddd is the truncated fraction of the day. The strong source 
GUSBAD 950203.097 not listed in the BATSE catalog is illustrated 
in Figure 6. The main properties listed in the catalog are as follows.
Besides equatorial (RA, dec) coordinates two other types of positions
are provided for the time of detection, namely azimuth and altitude,
and in an equatorial system anchored on BATSE. Four time bins are
listed, viz., $k_{\rm trig}$ to mark the detection (see Sec. 2.2), 
$k_{\rm top}$ for the peak of the time profile, $k_{\rm last}$ for 
the last detection above the photon flux limit, and $k_{\rm far}$ for 
the final detection in the simulation for $V/V_{\rm max}$ discussed 
in Sec. 5. Photon peak flux,
fluence and $V/V_{\rm max}$ are listed, as well as the brightest and 
second brightest illuminated detectors. The hardness ratios
ch(x)/(ch 2 + ch 3) are given for $x = 1-4$ as well as the spectral 
index derived from the counts in channels 2 and 3.
 
\section{Discussion}

The GUSBAD catalog lists 589 GRBs that are not in the BATSE catalog.
How many GRBs in the BATSE catalog are missing from the GUSBAD ctalog?
For this purpose, we have to limit the comparison to the 1144 GRBs in 
the BATSE catalog that have $C_{\rm max}/C_{\rm min} \ge 1.0$ on a 
timescale of 1024 ms. Among these, we find that the GUSBAD catalog 
is missing 156 BATSE sources.
There are several reasons why the catalogs have different contents.
One reason is that the backgrounds used in the two catalogs are
semi-independent. The background used for the BATSE catalog is covering
a time stretch of 17.408 s preceding the burst detection by a time
interval anywhere from 0.0 to 17.344 s. The first background stretch
used in the GUSBAD detections also covers 17.408 s but precedes the
burst detection by 20.0 s. Therefore, depending on circumstances,
the two backgrounds range from independent to mostly overlapping.
It is useful to remember that for independent backgrounds, the
probability that a source in one survey is detected in the other
is $\sim$50\% if the source is at the detection limit.
Another source of difference in catalog content has to do with
the search procedure. For the GUSBAD catalog, we
avoided high densities of triggers related to geographical
location, proximity to active sources (sun, CygX-1, etc), and poor data.
The on-board trigger for the BATSE sources at the 1024 ms time scale 
was disabled over regions of high magnetospheric activity, and 
following a burst event to allow transmission of data to the ground.

The positions of GRBs in the GUSBAD catalog were derived from an
algorithm employing counts from all eight BATSE detectors. The
internal mean errors of the positions are $\pm 7.5\degr$ 
(see Sec. 3), based on bursts of GRO J0422+32, which were 
typically near our detection limit. For 439 weak GRBs in the BATSE
catalog with $1.0<C_{\rm max}/C_{\rm min}<2.0$ the r.m.s. value
of the listed location errors is $\pm 6.8\degr$. 
A comparison of positions of GRBs common to the two catalogs
shows a r.m.s. dispersion of the difference of $\pm 9.8\degr$, 
but the interpretation of this value is difficult, since the 
GUSBAD and BATSE positions are not independent.
The positions given in the BATSE catalog were based on the six 
(originally four) detectors whose normals are closet to the source 
position. The move from four to six detectors led to some changes 
in position as large as $50\degr-100\degr$ \citep{mee96}. 

We originally defined in this work the duration of a GRB as the
total time span from trigger to last observation above the detection
limit, so $T = 1.024(k_{\rm last}-k_{\rm trig}+1)$ s. It is useful in
outlining the time interval in which information about positions,
spectral properties, etc. can be obtained.
As discussed in Sec. 5, when we reduce the burst amplitude in
the simulation leading to $V/V_{\rm max}$, the detection time may
move forward, depending on the time profile of the burst.
Similarly, the end of the reduced burst, may happen earlier and 
earlier. So $T$ will decrease as the burst amplitude is reduced. 
We find, in fact, that most bursts have at the simulated limit of 
detection a duration of one bin of 1.024 s only. This is to be 
expected: if there were two bins left above the detection limit, 
the amplitude could be reduced further resulting in only one bin. 
The variation of $T$ with simulated burst amplitude is so strong
that it makes it useless as an indicator of the physical duration 
of a GRB. This phenomenon is similar to the fluence duration 
bias in the BATSE catalog discussed by \citet{hak000}. The question
as to whether there is any definition of GRB duration that is robust 
and physically meaningful is beyond the scope of this article.

\acknowledgments

Following the declination of a grant proposal to the National Aeronautics
and Space Administration (NASA) in 1994, this work was carried out and 
completed without benefit of funds. This research made use of data 
obtained from HEASARC, provided by NASA's Goddard Space Flight Center.
It is a pleasure to thank J. Bonnell, J. Brainerd, D. Chakrabarty, 
V. Connaughton, M. Finger, G. Fishman, J. Gunn, A. Harmon, J. Higdon, 
J. Horack, M. McCollough, C. Meegan, G. Pendleton, T. Prince, C. Shrader,
B. Vaughan and K. Watanabe for assistance or information supplied during 
this project.

\begin{figure}
\includegraphics[angle=00,scale=.90]{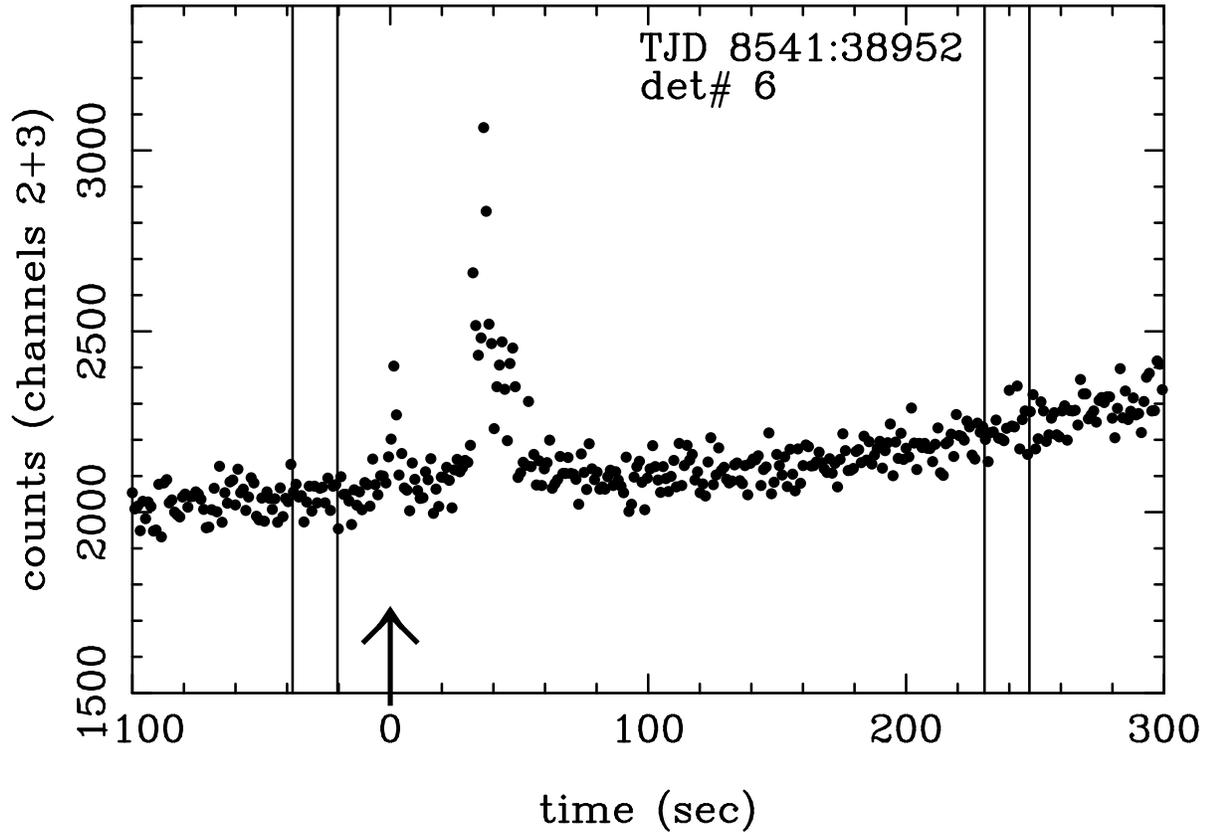}
\caption{In testing for the presence of a burst at $t=0.0$, the
background is linearly interpolated between windows centered at 
$-28.7$ s and $+239.1$ s. \label{fig1}}
\end{figure}

\clearpage

\begin{figure}
\includegraphics[angle=00,scale=.90]{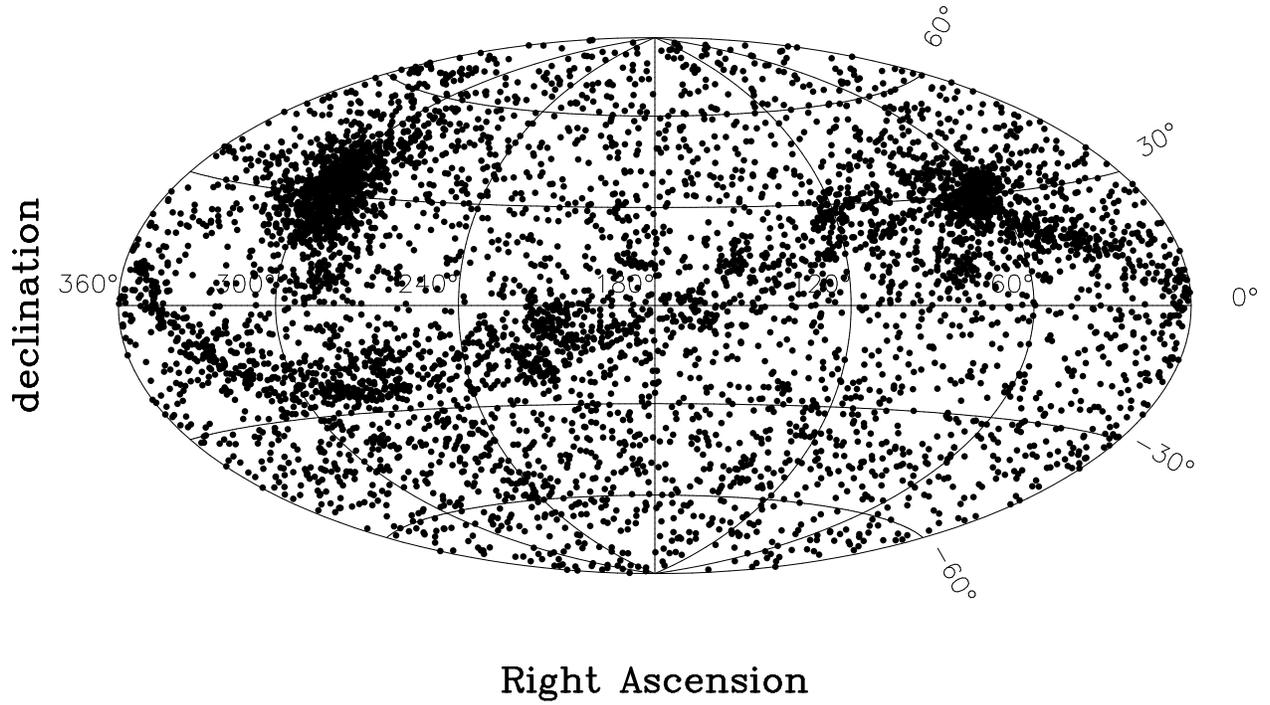}
\caption{Equatorial coordinates of 6236 trigger events. In the N.
hemisphere, CygX-1 and GRO J0422+32 are clearly seen. More than 1500
solar flares outline the ecliptic. The remaining triggers are
mostly composed of magnetospheric events and cosmic GRBs.\label{fig2}}
\end{figure}

\clearpage

\begin{figure}
\includegraphics[angle=-90,scale=.90]{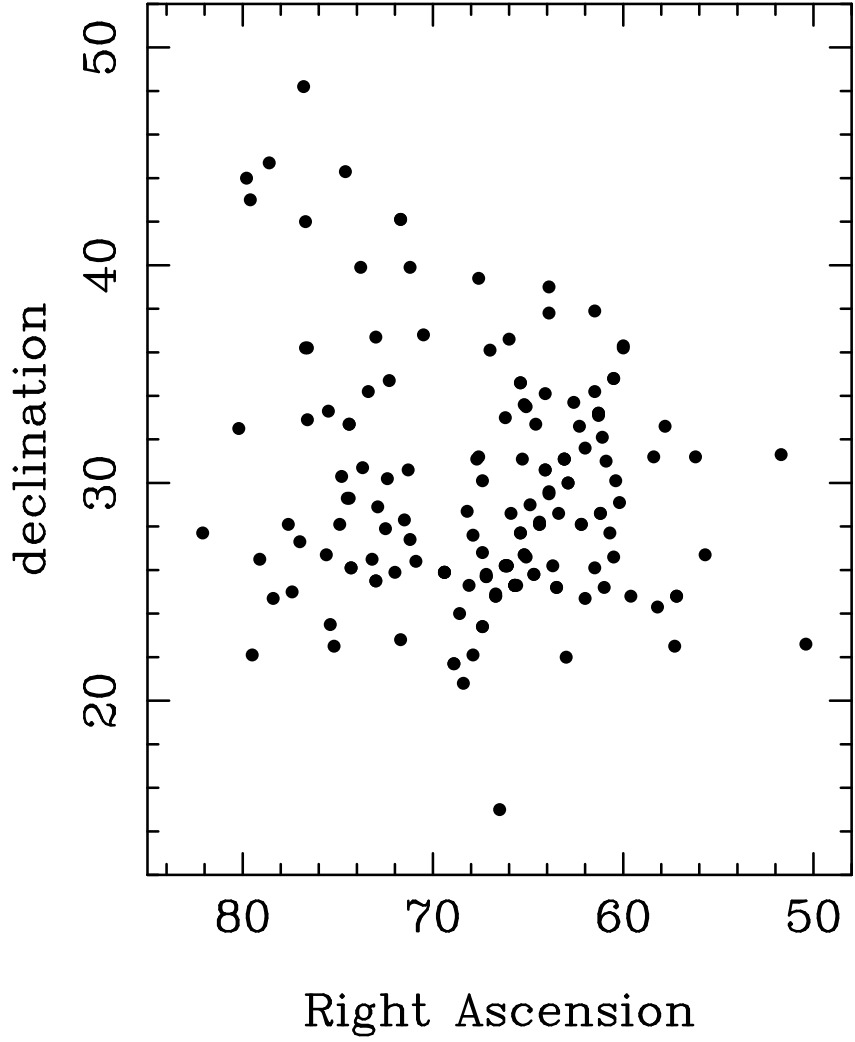}
\caption{Equatorial coordinates for 162 triggers of GRO J0422+32.
\label{fig3}}
\end{figure}

\clearpage

\begin{figure}
\epsscale{.80}
\includegraphics[angle=-90,scale=.80]{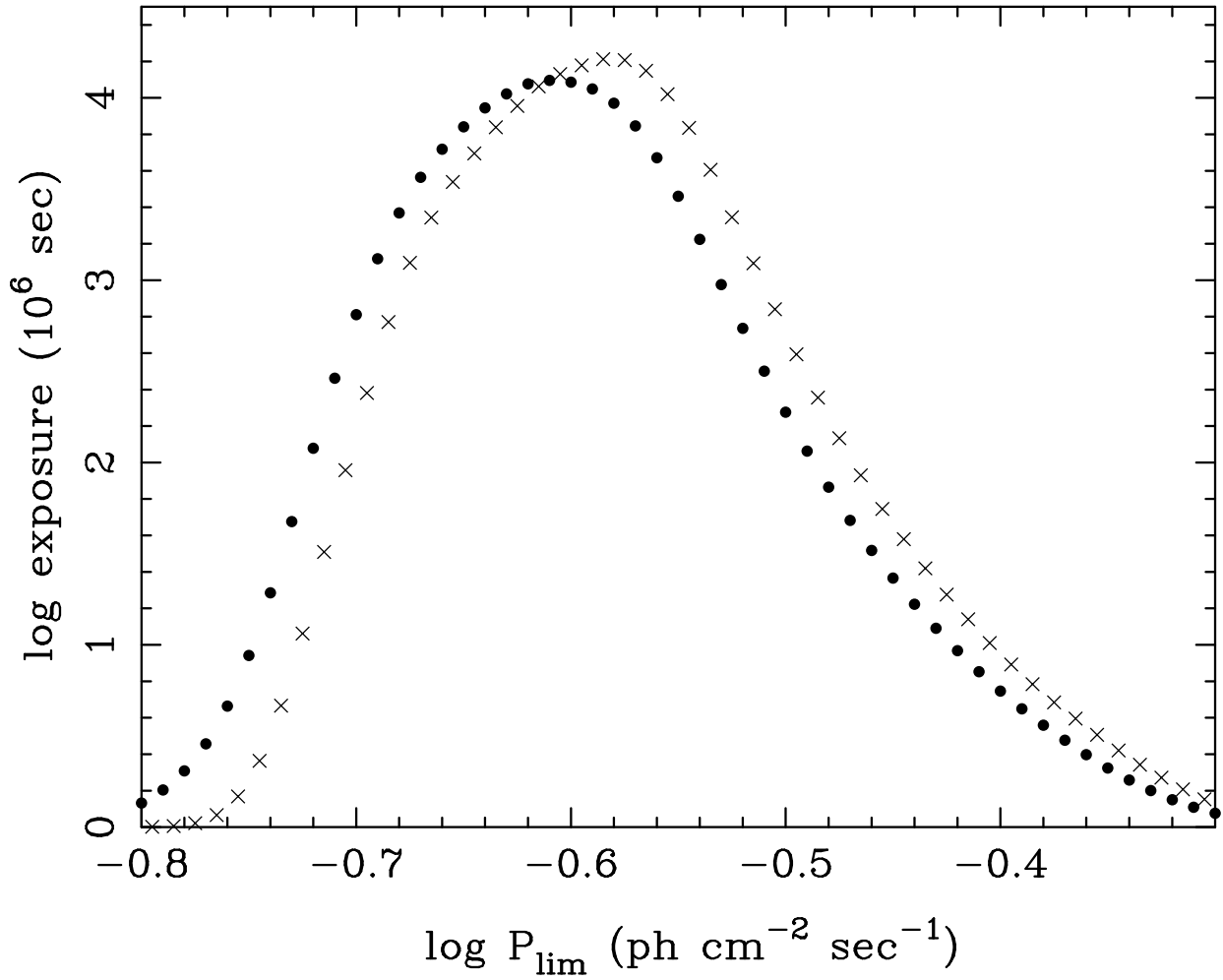}
\caption{Exposure as a function of limiting peak flux $P_{\rm lim}$. 
The crosses apply to direct radiation only; the dots include the effect 
of scattered radiation. The sum of the ordinates is the total exposure  
$1.0052 \times 10^8$ s.
\label{fig4}}
\end{figure}

\begin{figure}
\epsscale{.80}
\includegraphics[angle=-90,scale=.90]{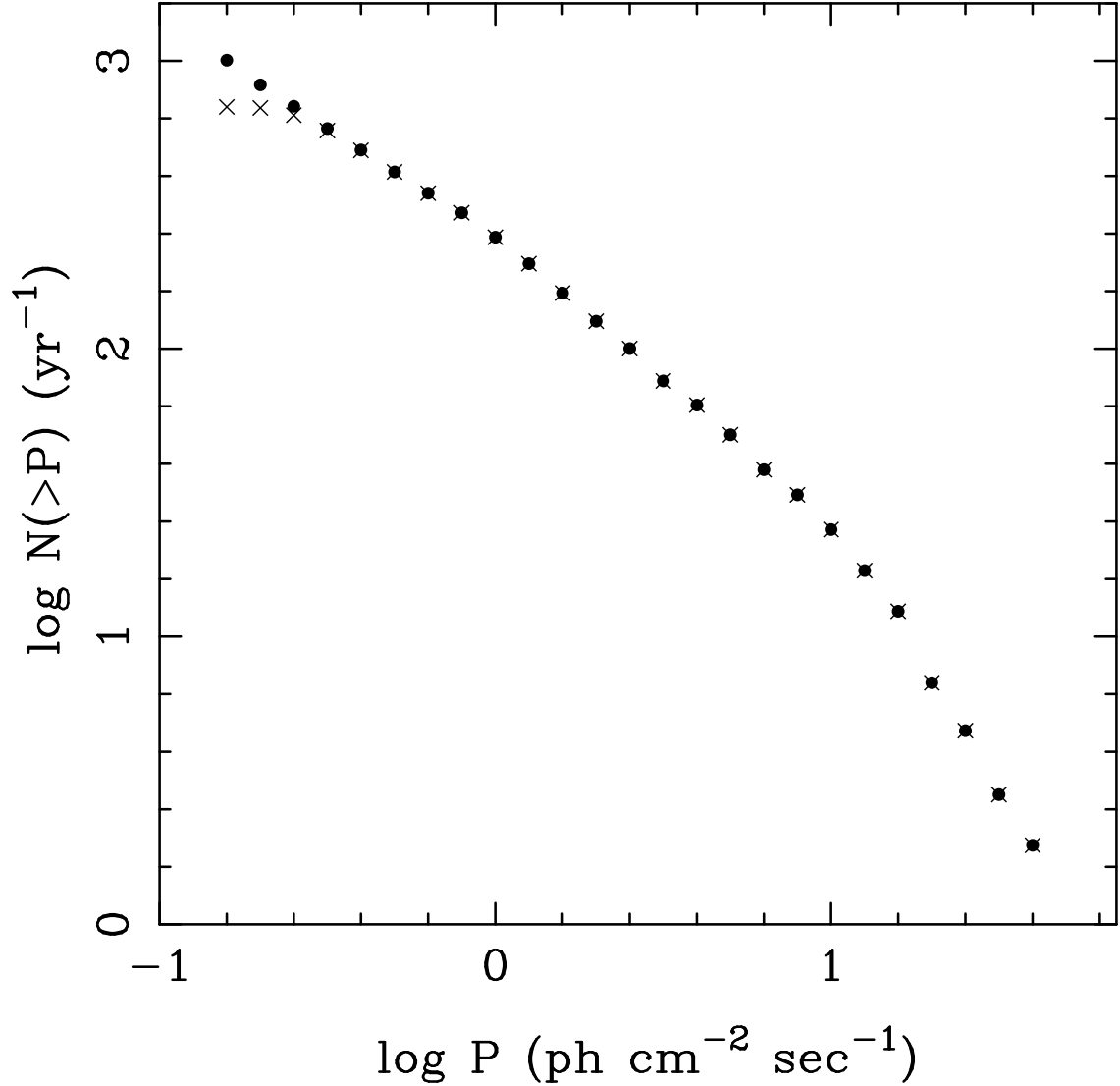}
\caption{Integral GRB source counts $N(>P)$ as a function of
peak flux $P$. Crosses are based on the raw source counts and the
total exposure time of 3.185 y; the dots account for the variation of
exposure time with $P$ (see Fig. 4). 
\label{fig5}}
\end{figure}

\begin{figure}
\epsscale{.80}
\includegraphics[angle=-90,scale=.80]{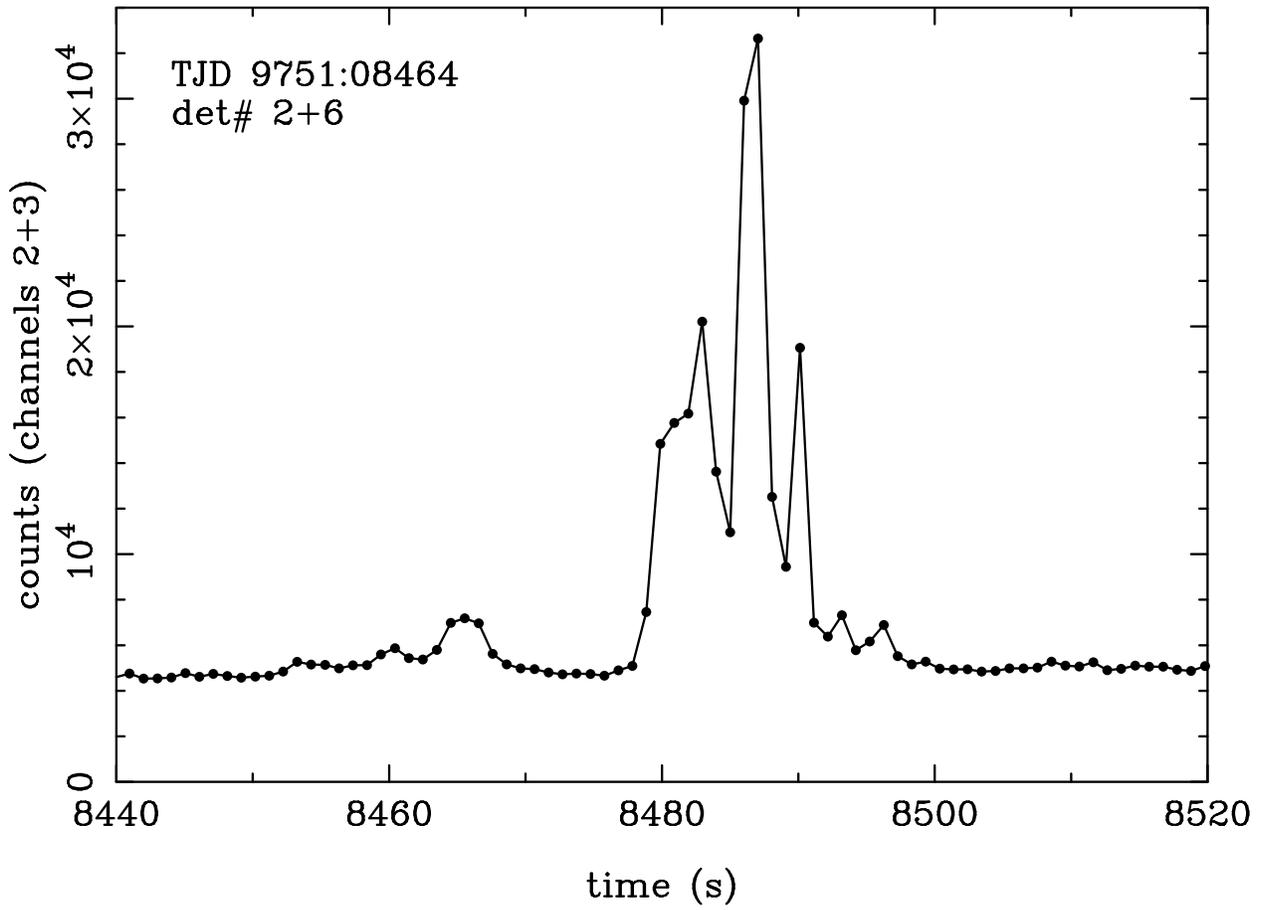}
\caption{GUSBAD 950203.097 is the strongest GUSBAD GRB not listed
in the BATSE catalog.
\label{fig6}}
\end{figure}




\clearpage

\begin{thebibliography}{}
\bibitem[Band et al.(1993)]{ban93} Band D. L. et al. 1993, \apj, 413, 281
\bibitem[Fishman et al.(1989)]{fis89} Fishman, G. J. et al. 1989,
    {\it GRO} Science Workshop Proc., W. N. Johnson,
    Greenbelt: NASA, 2-39
\bibitem[Fishman et al.(1992)]{fis92} Fishman, G. J., Meegan, C. A.,
    Wilson, R. B., Horack, J. M., Brock, M. N., Paciesas, W. S., 
    Pendleton, G. N. \& Kouveliotou, C. 1992, AIP Conf. Proc. 265,
    W. S. Paciesas \& G. J. Fishman, New York: AIP, 13
\bibitem[Fishman et al.(1994)]{fis94} Fishman, G. J. et al. 1994,
    \apjs, 92, 229
\bibitem[Hakkila et al.(2000)]{hak000} Hakkila, J., Meegan, C. A.,
    Pendleton, G. N., Mallozzi, R. S., Haglin, D. J. \& Roiger, R. J.
    2000, AIP Conf. Proc. 526, Gamma-Ray Bursts, R. M. Kippen, 
    R. S. Mallozzi \& G. J. Fishman, New York: AIP, 48
\bibitem[Higdon \& Lingenfelter(1996)]{hig96} Higdon, J. C. \&
    Lingenfelter, R. E. 1996, AIP Conf. Proc. 384, Gamma-Ray Bursts,
    C. Kouveliotou, M. F. Briggs \& G. J. Fishman,
    New York: AIP, 402
\bibitem[Horack et al.(1992)]{hor92} Horack, J. M., Fishman, G. J., 
    Meegan, C. A., Wilson, R. B. \& Paciesas, W. S., 1992, AIP Conf. 
    Proc. 265, W. S. Paciesas \& G. J. Fishman, New York: AIP, 373
\bibitem[Kommers et al.(1997)]{kom97} Kommers, J. M., Lewin, W. H. G.,
    Kouveliotou, C., van Paradijs, J., Pendleton, G. N., Meegan, C. A.
    \& Fishman, G. J. 1997, \apj, 491, 704
\bibitem[Meegan et al.(1996)]{mee96} Meegan, C. A. et al. 1996,
    AIP Conf. Proc. 384, C. Kouveliotou, M. F. Briggs \&
    G. J. Fishman, New York: AIP, 291
\bibitem[Schmidt(1999a)]{sch99a} Schmidt, M. 1999a, A\&AS, 138, 409
\bibitem[Schmidt(1999b)]{sch99b} Schmidt, M. 1999b, \apjl, 523, L117
\bibitem[Schmidt(2001)]{sch01} Schmidt, M. 2001, \apjl, 552, 36
\bibitem[Schmidt(2003)]{sch03} Schmidt, M. 2003, GUSBAD Catalog
\bibitem[Stern et al.(2001)]{ste01} Stern, B. E., Tikhomirova, Y.,
    Kompaneets, D., Svensson, R. \& Poutanen, J. 2001, \apj, 563, 80
    

\end{thebibliography}
\end{document}